\documentstyle[12pt,epsf]{article}

\begin{document}

\begin{flushright} 
CEBAF TH-96-05 \\
April 14, 1996
\end{flushright} 
\vspace{2cm}
\begin{center}
{\Large \bf Scaling Limit
of  Deeply Virtual Compton Scattering}
\end{center}
\begin{center}
{A.V. RADYUSHKIN\footnotemark }  \\
{\em Physics Department, Old Dominion University,}
\\{\em Norfolk, VA 23529, USA}
 \\ {\em and} \\
{\em Continuous Electron Beam Accelerator Facility,} \\
 {\em Newport News,VA 23606, USA}
\end{center}

\vspace{2cm}

\footnotetext{Also Laboratory of Theoretical Physics, 
JINR, Dubna, Russian Federation}

\begin{abstract}

I outline a perturbative QCD approach to the analysis
of the   deeply virtual 
Compton scattering  process $\gamma^* p \to \gamma p'$ 
in the  limit of  vanishing momentum transfer 
$t= (p' - p)^2$. The DVCS
amplitude in this limit exhibits   a 
scaling behavior  described by   
two-argument distributions
$F(x,y)$ which specify the fractions of the 
initial momentum $p$ and the momentum transfer 
$r \equiv p'-p$ carried by the constituents 
of the nucleon.
The kernel $R(x,y;\xi,\eta)$ governing the evolution
of the double distributions
$F(x,y)$ has a remarkable property:
it produces  the GLAPD evolution kernel $P(x/\xi)$ 
when  integrated  over $y$ and reduces  to the
Brodsky-Lepage evolution kernel $V(y,\eta)$ after the
$x$-integration. This property is used to 
construct the solution of the 
one-loop evolution equation for the flavor non-singlet 
part of the double quark distribution.  

\end{abstract}

\newpage

{\it 1. Introduction.} Recently,  X. Ji \cite{ji} suggested
to use the deeply virtual Compton scattering (DVCS) to 
get information about 
some parton distribution functions inaccessible 
in standard inclusive measurements.
He  considers     the
non-forward light-cone matrix  elements  
\begin{equation} 
\langle P-r/2  \, | \, \bar q (-{\lambda n/ 2}) \{ 1, \gamma_5\}
 \gamma^\mu q(\lambda n/2) \, | \, P +r/2\rangle,
\label{eq:ji}
\end{equation}
which appear in the lowest-order pQCD contribution to
the DVCS amplitude  ($r$ is 
 the momentum transfer and $n$ a light-like
4-vector) and parameterizes   them 
 using the functions   $H(x, r^2, r\cdot n),$ $ etc.,$
with $x$ being the Fourier conjugate variable to $\lambda$.
He observes that, in the $r \to 0$ limit,
the matrix element (\ref{eq:ji}) defines 
the usual  distribution functions
like $f(x)$, $ g_1(x)$ and proposes that the DVCS 
process can be used to get information
about  such  functions. 
Since the kinematics
of the DVCS  requires that $r \neq 0$, 
the $r^2 \equiv t \to 0$ limit can be accessed 
only by  extrapolating the 
small-$t$ data to $t=0$. 
The  limit $r \to 0$ 
looks  even more tricky  since $(r \cdot n) = 0$ 
formally corresponds to vanishing  of the
Bjorken variable $x_{Bj}$.
Anyway, as emphasized  by Ji \cite{ji}, the DVCS amplitude 
has a scaling   behavior  
in the region  of small $t$ and fixed $x_{Bj}$
which  makes it a very interesting object 
on its own ground.

In  this letter, I  briefly describe   an
alternative  pQCD   formalism   for the
analysis  of the DVCS  amplitude in the   
limit  when $t \to 0$ and $x_{Bj}$ is fixed.
My main point is that,
to construct a consistent pQCD picture for 
the scaling limit  of   DVCS, 
one should treat the initial momentum $p$ and 
the momentum transfer $r$ on 
equal footing by  introducing 
double distributions
$F(x,y)$, which specify the fractions of  $p$ 
and  $r$, $resp.,$ 
carried by the constituents 
of the nucleon\footnote{I found no
advantage in introducing the average nucleon 
momentum $P=(p+p')/2$ like in eq. (\ref{eq:ji})}. 
These distributions have hybrid properties:
they look like distribution functions 
with respect to $x$ and like  distribution amplitudes
with respect to $y$.
Writing the matrix element of a composite operator 
in terms of the  double distributions
is the first step in developing  the pQCD parton picture
for the DVCS.  The next step is to take into account 
the logarithmic scaling violation. 
To this end, I  write down  the evolution
equation for the simplest case of the  flavor non-singlet 
part of the double quark distribution
$F^{NS}(x,y;\mu)$.  As one may expect
from the preceding description, the relevant evolution
kernel $R(x,y;\xi,\eta)$ has a remarkable property:
it produces  the GLAPD evolution 
kernel $P^{NS}(x/\xi)$ \cite{gl,ap,d}
when  integrated  over $y$,  while 
integrating  $R(x,y;\xi,\eta)$ over $x$
gives   the expression coinciding with the 
Brodsky-Lepage evolution kernel $V(y,\eta)$  \cite{bl} 
for the pion distribution amplitude\footnote{Originally,
the evolution equation for the pion 
distribution amplitude in QCD
was derived and solved in ref.\cite{tmf},
where the anomalous dimension matrix 
$Z_{nk}$ was used instead of $V(x,y)$ (see also \cite{phlet}).}.
Using  these properties of the kernel, 
I construct the solution of the 
one-loop evolution equation for the flavor non-singlet 
part of the double  distribution. 
I also  discuss the  infrared sensitivity
of the DVCS amplitude  due to the presence 
of the light-like momentum $q$ and 
its implications for the structure of the 
simplest radiative and higher-twist corrections
to the leading-twist result.

{\it 2. Double distributions. } 
The  kinematics of the 
amplitude of the process $\gamma^* p \to \gamma  p'$ can be 
most conveniently specified  
by the initial nucleon momentum $p$,
the   momentum  transfer $r=p-p'$ and  the 
momentum $q$  of the final photon. 
Since $q^2 =0$, it is natural to use $q$ 
as one of the basic light-cone (Sudakov) 4-vectors.
In  the scaling limit, the invariant momentum transfer 
$t \equiv r^2$ and the square of the proton mass $m_p^2=p^2$ 
can be neglected compared to
the  virtuality $-Q^2 \equiv (q-r)^2$ of the initial 
photon and the energy invariant $p \cdot q \equiv m_p \nu$. 
Thus,  we set  $p^2=0$ and $r^2 = 0$,
and use $p$ as another basic light-cone  4-vector.
Furthermore,  in this limit, the  requirement
$p'^2 \equiv (p+r)^2=p^2$  reduces to the condition 
$p\cdot r = 0$  which can be satisfied only
if the  two lightlike momenta
$p$ and $r$ are proportional to each other: $r= \zeta  p$,
where $\zeta$ coincides with 
the Bjorken variable $\zeta = x_{Bj} \equiv Q^2/2(p \cdot q)$ 
which satisfies the constraint  $0 \leq x_{Bj}  \leq 1$.

\begin{figure}[ht]
\mbox{
   \epsfxsize=12cm
 \epsfysize=5cm
 \hspace{2cm}  
 \epsffile{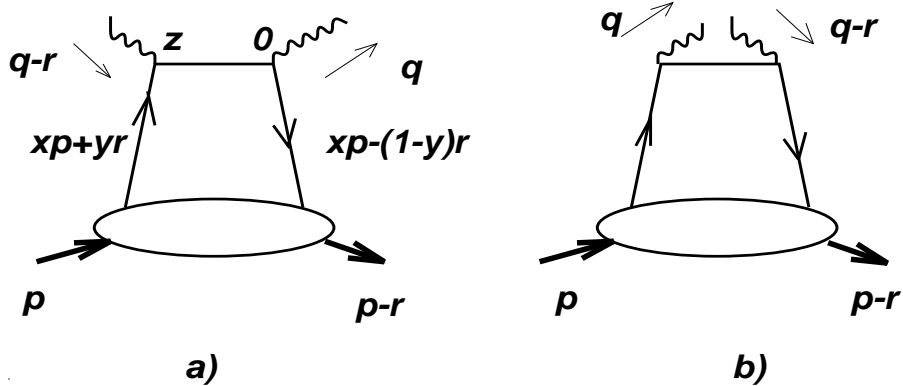}  }
{\caption{\label{fig:1}
Handbag diagrams contributing 
 into the DVCS amplitude. The lower blob corresponds 
to  double quark distributions $F(x,y), G(x,y)$.
   }
}
\end{figure}

The leading contribution in the  large-$Q^2$, fixed-$x_{Bj} $,
$t=0$  limit
is given by the handbag
diagrams shown on  Fig.\ref{fig:1}, in which  the long-distance 
dynamics is described by  matrix elements like 
\begin{equation} 
\langle p -r  \,  | \,   \bar \psi_a(0) \gamma_{\mu} 
E(0,z;A) \psi_a(z) \, | \,p  \rangle  \ \ \ {\rm and}  \ \ \ 
\langle p -r  \, | \, \bar \psi_a(0) \gamma_5 \gamma_{\mu} 
E(0,z;A) \psi_a(z) \, | \,p \rangle , 
\label{eq:my}
\end{equation}
where, for Fig.\ref{fig:1}$a$, $z$ is the coordinate 
of the virtual  photon vertex and $E(0,z;A)$ is the usual
$P$-exponential of the gluonic $A$-field  
along the straight line  connecting 0 and $z$.

Though the momenta
$p$ and $r$ are proportional to each other $r = \zeta  p$,
to construct an adequate QCD parton picture,
one should make a clear distinction between them.
The basic  reason 
is  that $p$ and $r$  specify 
the momentum flow in two different   channels.
For $r=0$, the net momentum flows only in the $s$-channel 
and the total momentum entering into 
the  composite operator vertex is zero. 
In this case, 
the matrix element coincides with  the 
standard   distribution function.
The  partons  entering the composite vertex 
then carry  the   fractions $x_i$ 
of the initial proton momentum ($-1 < x_i <1$). 
When  $x $ is  negative, the parton is  interpreted
as belonging to the final state and  $x_i$ is redefined
to  secure that  
the integral always  runs over  the segment  $0\leq x\leq 1$.
In this parton picture, the spectators take the 
remaining momentum  $(1-x)p$.
On the other hand, if  the 
total momentum flowing through the 
composite vertex is $r$, 
the matrix element has the structure
of the distribution amplitude in which 
the momentum  $r$  
splits into the fractions $yr$  and 
$(1-y)r \equiv \bar y r$ carried by the 
quark fields  attached to  that vertex. 
In a combined situation, when both $p$ and $r$
are nonzero,  the initial quark  has the momentum 
$xp +y r$, while the final one  carries the momentum 
$xp - \bar y r$.
In more formal terms, this corresponds  
to the following parameterization 
of the light-cone matrix elements 
\begin{eqnarray} 
&& \langle p-r\, | \, \bar \psi_a(0) \hat z 
E(0,z;A)  \psi_a(z) \, | \, p \rangle |_{z^2=0} 
 =  \bar u(p-r)  \hat z 
u(p)  \int_0^1   \int_0^1  \, 
  \left ( e^{-ix(pz)-iy(r z)}F_a(x,y) \right.
\nonumber \\
&& \hspace{4cm}  - \left.  e^{ix(pz)-i\bar y(r z)}F_{\bar a}(x,y)
\right )
 \, \theta( x+y \leq 1) dy \, dx ,
\label{eq:vec}
\end{eqnarray} 
\begin{eqnarray} 
&& \langle p-r\, | \, \bar \psi_a(0) \gamma_5 \hat z 
E(0,z;A)  \psi_a(z) \, | \, p \rangle |_{z^2=0}  
 =  \bar u(p-r)  \gamma_5 \hat z 
u(p)  \int_0^1   \int_0^1  \, 
  \left ( e^{-ix(pz)-iy(r z)}G_a(x,y) \right.
\nonumber \\
&& \hspace{4cm} + \left.  e^{ix(pz)-i\bar y(r z)}
G_{\bar a}(x,y) \right ) \, 
\theta( x+y \leq 1) dy \, dx ,
\label{eq:ax}
\end{eqnarray} 
where $\hat z \equiv \gamma_{\mu} z^{\mu}$ 
and $\bar u(p-r), u(p)$ are the Dirac spinors for the nucleon.
 
Though  we arrived at the matrix elements  (\ref{eq:vec}), (\ref{eq:ax})
in the context of the scaling limit of the DVCS amplitude,
they accumulate a  process-independent information and, hence,
have a quite general nature.
The coefficient of proportionality between 
$p$ and $r$ is then just the  parameter  characterizing
the ``asymmetry'' of the matrix elements.
The fact that, in our case, $\zeta$ coincides with 
the Bjorken variable is  specific  for 
the DVCS amplitude.
The most non-trivial feature  implied by
the representation (\ref{eq:ax}) is the absence 
of the $\zeta$-dependence in 
the double distributions $F_a(x,y)$ and $G_a(x,y)$.
Using the methods  developed in ref.\cite{spectral},
one  can  prove that this property and the 
spectral constraints $x \geq 0$,
$y \geq 0$, $x+y \leq 1$ 
 hold for  any Feynman diagram.
As a result,  both the initial active quark 
and the spectators  carry  positive 
fractions of the light-cone ``plus'' momentum $p$:
$x+\zeta y$ for the active quark and 
$\bar x- \zeta y  = (1-x-y) +(1-\zeta )y$ 
for   the spectators. 
Note, however, that the fraction of the initial momentum  $p$
carried by the 
quark going out of the composite  vertex is given by 
$x - \bar y \zeta $ and it may take  both 
positive and negative values.
At first sight, this result contradicts the intuition  
based on the infinite momentum frame picture.
Recall,  however, that all the fractions are positive only when  
the ``plus'' component
of the momentum transfer $r$ vanishes,
which is not the case here, since $r$ has $only$ 
the ``plus'' component.

Taking the  limit $r =0$ gives  the matrix
element defining the parton distribution functions $f_a(x)$,
$f_{\bar a}(x)$ (or $g_a(x)$,
$g_{\bar a}(x)$).
This  observation results   in the following reduction formulas
 for  the double distributions $F(x,y), G(x,y)$:
\begin{equation}
\int_0^{1-x} \, F_a(x,y)\, dy=  f_a(x) \  \ ,  \  \ 
\int_0^{1-x} \, G_a(x,y)\, dy=  g_a(x).
\label{eq:redf}
\end{equation}

{\it 3. Leading-order contribution.}
Using the parameterization for the matrix elements
given above, we get a parton-type representation
for the handbag contributions to the DVCS amplitude:
\begin{eqnarray} 
T^{\mu \nu} (p,q,r) =  
\left (g^{\mu \nu} -\frac1{p \cdot q } 
(p^{\mu}q^{\nu} +p^{\nu}q^{\mu}) \right ) \, 
\sum_a 
e_a^2\, \sqrt{1- \zeta} \ ( T_V^a(\zeta ) + T_V^{\bar a}(\zeta ) ) \nonumber \\
+i \epsilon^{\mu \nu \alpha \beta} p_{\alpha} q_{\beta} 
\,  
\frac{ \bar u(p') \gamma_5 
\hat q  u(p) }{2\, (p\cdot q)^2} \, \sum_a 
\, e_a^2\,   ( T_A^a(\zeta ) - T_A^{\bar a}(\zeta ) ) \, ,
\label{eq:hbag}
\end{eqnarray} 
where $\hat q \equiv  \gamma_{\mu}q^{\mu}$,  the factor 
$\sqrt{1- \zeta}$ comes from $\bar u (p') = \sqrt{1- \zeta} \, \bar u(p)$
and 
$T_V^a(\zeta )$, $T_A^a(\zeta )$ are the 
invariant amplitudes 
depending on  the scaling variable $\zeta $:
\begin{eqnarray} 
T_V^a(\zeta ) =  \int_0^{1}  dx \, \int_0^{1-x} \left
 ( \frac1{x-\zeta \bar y+i\epsilon}
+ \frac1{x+\zeta  y} \right ) F_a(x,y) \,  dy , 
\label{eq:tv} \\
T_A^a(\zeta )  = 
\int_0^1  dx \, \int_0^{1-x} \left ( \frac1{x-\zeta \bar y+i\epsilon}
- \frac1{x+\zeta  y} \right ) G_a(x,y) \, dy.
\label{eq:ta}
\end{eqnarray} 
The terms  containing $1/(x-\zeta \bar y+i\epsilon)$ 
generate the imaginary part:
\begin{eqnarray} 
 \frac1{\pi}\, {\rm Im} \, T_V^a(\zeta ) = 
 \int_0^1 \int_0^{1}  \delta( x - \zeta \bar y) F_a(x,y) \, 
\theta (x+y \leq 1) \, dx \, dy
\nonumber \\ 
=  
\, \frac1{\zeta } \int_0^{\zeta } F_a(x, 1-x/\zeta ) dx =  
 \int_0^1
F_a( \bar y \zeta ,  y ) \, d  y , 
\label{eq:imtv}
\end{eqnarray} 
with a similar expression for $ {\rm Im} \, T_A^a(\zeta )$.
Because of the
integration  remaining in eq.(\ref{eq:imtv}), the  relation between 
${\rm Im} \, T(\zeta )$ and the double distributions $F_a(x,y)$
is not as direct as in the case of forward virtual 
Compton amplitude, the imaginary part of which 
is just given by distribution functions $f_a(\zeta)$. 
Note, that the $y$-integral in eq.(\ref{eq:imtv}) 
 is different  from that in the reduction formula 
(\ref{eq:redf}), $i.e.,$ though 
\begin{equation}
\Phi_a(\zeta) \equiv \int_0^1
F_a( \bar y \zeta ,  y ) \, d  y 
\end{equation}
is a function of the Bjorken  variable $\zeta$, it does 
not coincide with $f_a(\zeta)$.

To get the real part of the $1/(x-\zeta \bar y+i\epsilon)$ terms,
one should use the principal value prescription,
$i.e.,$ ${\rm Re} \, T(\zeta )$ is related to 
$F_a(x,y)$ through two integrations.

{\it 4. Evolution equation.} 
In QCD, the limit $z^2 \to 0$
for  the matrix elements  in eq.(\ref{eq:my})
is singular.  As a result, in perturbation theory,
the distribution  $F(x,y)$ contains 
logarithmic ultraviolet divergences 
which can be removed in a standard way by applying the
$R$-operation characterized by some subtraction scale
$\mu$:  $F(x,y)\to F(x,y;\mu)$.
The $\mu$-dependence of  $ F(x,y;\mu)$  is governed by the
evolution equation
\begin{equation}
\left ( \mu \frac{\partial}{\partial \mu} +\beta(g)
 \frac{\partial}{\partial g} \right ) F(x,y;\mu) =
\int_0^1 d \xi \int_0^1  R(x,y; \xi, \eta;g) 
F( \xi, \eta;\mu) d \eta.
\label{eq:nfwdev} 
\end{equation}
Since the integration over $y$ converts $F(x,y)$ 
into the parton distribution function $f(x)$,
whose evolution is governed by the GLAP equation
\begin{equation}
\left ( \mu \frac{\partial}{\partial \mu} +\beta(g)
 \frac{\partial}{\partial g} \right ) f(x;\mu) =
\int_x^1 \frac{d\xi}{\xi} P(x / \xi;g) 
f( \xi;\mu) d \xi, 
\label{eq:glap}
\end{equation}
the kernel $R(x,y; \xi, \eta;g)$ must have  the property
\begin{equation}
\int_0^ {1-x}  R(x,y; \xi, \eta;g) d y =
\frac1{\xi} P(x/\xi).
\label{eq:rtop}
\end{equation}
For a similar reason, integrating $R(x,y; \xi, \eta;g)$
over $x$ one should get  the Brodsky-Lepage  kernel:
\begin{equation}
\int_0^{1-y}   R(x,y; \xi, \eta;g) d x = V(y,\eta;g).
\label{eq:rtov}
\end{equation}
 Explicit one-loop calculations give the following result 
 for the $qq$-component of the kernel:
\begin{eqnarray}
&& \hspace{-1.5cm} R_{qq}(x,y;\xi, \eta;g) = 
\frac{\alpha_s}{\pi} C_F \frac1{\xi}
\left \{  \theta  (0 \leq x/\xi \leq 
{\rm min} \{ y/\eta, \bar y / \bar \eta \} ) -
\frac1{2} \delta(1-x/\xi) \delta(y-\eta) \right. \label{eq:rkernel}
 \\
 && \hspace{-1.5cm}  +\left.   
\frac{\theta (0 \leq x/\xi \leq 1) x/\xi}{ (1-x/\xi)} 
\left [ \frac1{\eta}\delta(x/\xi - y/\eta) + 
\frac1{\bar \eta} \delta(x/\xi - \bar y/ \bar \eta) \right]
-2\delta(1-x/\xi) \delta(y-\eta)
\int_0^1 \frac{z}{1-z} \, dz \right \}.
\nonumber
\end{eqnarray}
Here the last (formally divergent) term, as usual,
provides the regularization for the singularities 
of the kernel for  $x=\xi$ (or $y= \eta$).
Note that, in the Feynman gauge, the first line of eq.(\ref{eq:rkernel})
corresponds to operators with ordinary derivatives $\partial^{\nu}$
while the second one results from the 
$\partial^{\nu} \to D^{\nu} =\partial^{\nu} -igA^{\nu} $ 
change. It is easy to verify that the kernel 
$R_{qq}(x,y;\xi, \eta;g)$ has the property that 
$x+y \leq 1$ if $\xi+ \eta \leq 1$. 
Using our  expression for $R_{qq}(x,y;\xi, \eta;g)$ and explicit
forms of the $P_{qq}(x/\xi)$ and  $V(y, \eta)$ kernels
\begin{eqnarray}
&&\lefteqn{ P_{qq}(z) = \frac{\alpha_s}{\pi} 
C_F \left (\frac{1+z^2}{1-z} \right )_+ ,} 
\label{eq:pkernel} \\
&& V(y, \eta)  = \frac{\alpha_s}{\pi} C_F 
\left \{  \left (\frac{y}{\eta} \right )
\left [1+ \frac1{ \eta -y} \right ]
\theta(y \leq \eta) + 
\left (\frac{\bar y}{\bar \eta} \right )
\left [1+ \frac1{y- \eta} \right ]
\theta(y \geq \eta) 
  \right \}_+ 
\label{eq:vkernel} 
\end{eqnarray}
(where ``+'' denotes the standard ``plus''
regularization \cite{ap}), one can  check 
that $R_{qq}(x,y; \xi, \eta;g)$ satisfies the reduction formulas
(\ref{eq:rtop}) and (\ref{eq:rtov}).  

Note also that the flavor non-singlet 
light-cone operator ${\cal O}^{NS}(z,0)\equiv 
\bar \psi (z) \lambda^a \hat z \psi (0) $ 
satisfies the Balitsky-Braun evolution equation \cite{bb} 
\begin{equation}
\left ( \mu \frac{\partial}{\partial \mu} +\beta(g)
 \frac{\partial}{\partial g} \right )
{\cal O}^{NS}(z,0)  =
\int_0^1 d u \int_0^{u}   
K(u,v ) {\cal O}^{NS} (uz,vz)\, d v . 
\end{equation}
This means that our kernel $R_{qq}(x,y; \xi, \eta;g)$  should be 
  related to  the $K$-kernel by
\begin{equation}
 R_{qq}(x,y; \xi, \eta;g) = \frac1{\xi} 
K( \bar y + \eta x/\xi, \bar y - \bar \eta x/\xi).
\label{eq:bbtor}
\end{equation}
Indeed, taking  the explicit form  of $K(u,v )$ from ref. 
\cite{bb}\footnote{In the definition adopted in ref.
\cite{bb}  $K(u,v )$ 
has the opposite sign.}
\begin{equation}
K(u,v ) = \frac{\alpha_s}{\pi} C_F \left 
(1 + \delta(\bar u) [\bar v/v]_+  + 
\delta(v) [u/\bar u]_+ - \frac1{2} \delta(\bar u)\delta(v) \right )
\end{equation}
and combining it with eq.(\ref{eq:bbtor}),  one 
immediately obtains our expression (\ref{eq:rkernel}) for
$R_{qq}(x,y; \xi, \eta;g)$.

To solve  the evolution equation,  
I propose to combine
the standard methods used to find  solutions of  the underlying
GLAP and Brodsky-Lepage evolution equations.
Here, we  will consider only the   simplest 
case, $i.e.,$ the evolution equation for the
 flavor-nonsinglet component in  which  $qg$, $gq$ and $gg$ 
kernels do not contribute.  
 Recall first that, to solve  the GLAP equation, 
one should consider the moments with respect to $x$.
Integrating $x^n R_{qq}(x,y; \xi, \eta;g)$ over $x$ and utilizing 
the property 
 $R_{qq}(x,y; \xi, \eta;g) = R_{qq}(x/\xi,y; 1, \eta;g)/\xi$,
we get  
\begin{equation}
\left ( \mu \frac{\partial}{\partial \mu} +\beta(g)
 \frac{\partial}{\partial g} \right ) F_n(y;\mu) =
\int_0^1  R_n (y,\eta;g) F_n(\eta;\mu) d \eta \,  , 
\label{eq:fnev}
\end{equation}
where  $F_n(y;\mu)$ is the $n$th $x$-moment of $F(x,y;\mu)$
\begin{equation}
F_n(y;\mu) = \int_0^{1} x^n  F(x,y;\mu) dx
\label{eq:fnmom}
\end{equation}
and the kernel $R_n (y,\eta;g)$  is given by
\begin{eqnarray}
R_n (y,\eta;g) = 
\frac{\alpha_s}{\pi} C_F 
\left \{  \left (\frac{y}{\eta} \right )^{n+1} 
\left [\frac1{n+1}+ \frac1{ \eta -y} \right ]
\theta(y \leq \eta) + 
\left (\frac{\bar y}{\bar \eta} \right )^{n+1} 
\left [\frac1{n+1} + \frac1{y- \eta} \right ]
\theta(y \geq \eta) \right. \nonumber \\ \hspace{-3cm}
\left. - \frac1{2} \delta(y-\eta) -  \delta(y-\eta)
\int_0^1 \frac{z}{1-z} \, dz  \right \}.
\label{eq:rnkernel}
\end{eqnarray}
It is straightforward to check 
that  $R_n (y,\eta;g)$ has the property
$$R_n (y,\eta;g) w_n(\eta) =
R_n (\eta,y;g) w_n(y) ,$$ where $w_n(y)= (y \bar y)^{n+1}$. 
Hence, the  eigenfunctions of $R_n (y,\eta;g)$ are orthogonal with
the weight $w_n(y)= (y \bar y)^{n+1}$, $i.e.,$ 
they are proportional to the Gegenbauer polynomials
$C^{n+3/2}_k(y-\bar y)$ (cf.\cite{bl,mikhrad}).
Now, we can write  the general solution of the evolution
equation
\begin{equation} 
F_n(y;\mu) = (y \bar y)^{n+1} \sum_{k=0}^{\infty} A_{nk}
C^{n+3/2}_k(y-\bar y) \left [\log (\mu /\Lambda) \right]^
{-\gamma^{(n)}_k/\beta_0},
\label{eq:rnsol}
\end{equation}
where $\beta_0 = 11-\frac2{3}N_f$ is the 
lowest coefficient of the QCD $\beta$-function and 
the anomalous dimensions $\gamma^{(n)}_k$ 
 are related to   the eigenvalues of the kernel $ R_n (y,\eta;g)$:
\begin{equation}
\gamma^{(n)}_k= C_F \left [ \frac1{2} - 
\frac1{(n+k+1)(n+k+2)} +2 \sum_{j=2}^{k+n+1} \frac1{j}
\right ].
\label{eq:ads}
\end{equation}
They coincide with the standard non-singlet 
anomalous dimensions $\gamma_{N}$ \cite{gw,gp}:  
$\gamma^{(n)}_k = \gamma_{n+k+1}$. 
Note, that $\gamma_0^{(0)}=0$, while  all other anomalous  dimensions 
are positive. Hence,  in the formal $\mu \to \infty$
limit, we have  $F_0(y, \mu \to \infty) \sim y \bar y$ and 
$F_n(y, \mu \to \infty) =0$  for all $n \geq 1$.
This means that 
$$F(x,y; \mu \to \infty)  \sim \delta(x) y \bar y,$$
$i.e.,$ in each of its variables, the limiting function 
 $F(x,y; \mu \to \infty)$  
acquires the characteristic asymptotic form dictated by
the nature of the variable:
$\delta(x)$ is specific for the distribution functions \cite{gw,gp},
while  the $y \bar y$-form  is  
the asymptotic shape   for the lowest-twist two-body 
distribution amplitudes \cite{tmf,bl}.

{\it 5. Infrared sensitivity and 
hard gluon exchange  corrections.} The structure 
of the leading term of the DVCS amplitude $T(\zeta , Q^2)$ 
is very   similar to that of the forward virtual Compton 
scattering amplitude $T_f(\omega,Q^2)$ 
which is the starting point of the 
DIS analysis. The  major difference  between the two
amplitudes is that
one of the photons in the DVCS amplitude   is real.
As a result, in higher orders, the coefficient functions 
$C(\zeta ,Q^2,q^2)$ of the formal
operator product expansion of two electromagnetic currents
may be singular  (non-analytic) in the limit $q^2 \to 0$.
These singularities  are related to the possibility 
of a  long-distance propagation
in the $q$-channel. In fact, the long-distance  sensitivity
simply means that the relevant contribution 
is non-calculable in perturbation theory, and one should 
describe/parameterize it by introducing the 
distribution amplitude for the real photon.
 
\begin{figure}[ht]
\mbox{
   \epsfxsize=12cm
 \epsfysize=5cm
 \hspace{2cm}  
 \epsffile{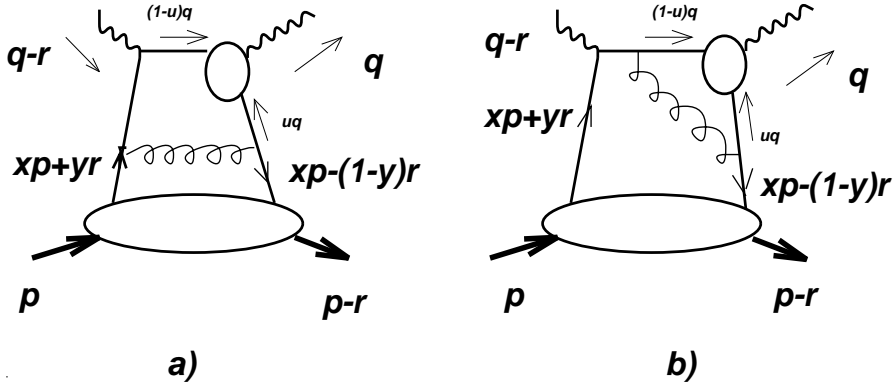}  }
{\caption{\label{fig:2}
Simplest hard gluon exchange  contributions
to the DVCS amplitude.  The upper
 blob corresponds  to the photon distribution
amplitude $\varphi_{\gamma}(u)$ and the lower one 
to  double quark distributions $F(x,y), G(x,y)$.
   }
}
\end{figure}

Two simplest contributions of this type are shown in fig.\ref{fig:2}.
They are analogous to the hard gluon exchange diagrams 
for the pion form factor.  The basic common feature is that
the large virtuality flow bypasses the real photon vertex,
and the relevant contribution factorizes into 
the product of the hard scattering (short-distance) 
amplitude formed by 
quark and gluon propagators and two long-distance parts.
In our case, one long-distance part is given by the 
double distribution $F(x,y)$ (or $G(x,y)$) and
the other by the  photon distribution amplitude  
$\varphi_{\gamma}(u)$.
The latter can be understood as  the probability amplitude 
to obtain   the photon with momentum  $q$ 
from a collinear $\bar qq $ pair, with quarks 
carrying the momenta $u q$ and $(1 - u) q$.
The $u^n$-moments  of the lowest-twist
function $\varphi_{\gamma}(u)$
are related to the difference between the 
``exact'' and perturbative versions of the correlator
\begin{equation}
\Pi_{\mu \nu \nu_1 \ldots \nu_n} (q) \equiv
\int e^{iqz} \langle 0 | T \{ J_{\mu}(z) \  \bar \psi_a(0) \gamma_{\nu}
D_{\nu_1}  \ldots D_{\nu_n} \psi_a(0) \} | 0 \rangle d^4 z
\end{equation} 
of the electromagnetic current $J_{\mu}(z)$ 
with a composite operator containing $n$ covariant derivatives.
The perturbative amplitude 
for the $\bar q q \to \gamma$ transition is taken into account in the
pQCD radiative correction 
to the coefficient function ($cf.$ \cite{rr}).
By a simple counting of the propagators,
one  can easily find out  that the contribution 
of Fig.\ref{fig:2} behaves like $\alpha_s/Q^2$ in the large-$Q^2$ limit. 
Furthermore, due  to the EM current conservation, this 
correlator is proportional to the tensor structure
$g_{\mu \nu} q^2 - q_{\mu} q_{\nu}$. 
Since, for a real photon, $q^2 =0$ and $(q \epsilon) =0$  
($\epsilon^{\mu}$ being the photon's polarization),
  the relevant contribution  vanishes.
Hence, the hard gluon exchange corrections to the DVCS 
amplitude  are rather strongly suppressed.

At the leading-power level $O(Q^0)$, 
the absence of the non-analytic $\ln(q^2)$-terms 
in the one-loop  coefficient function 
contribution for the $A$-part of the amplitude 
can be read off from existing results for
the $\alpha_s$ correction to the $\gamma \gamma^* \to \pi^0$
amplitude \cite{AuCh81,braaten,kmr}.

{\it 6. Conclusions.}  In this paper, I 
formulated the basics of the pQCD approach for studying  
the deeply virtual Compton scattering 
amplitude in the scaling limit.  
The essential point is that information about 
the long-distance dynamics 
in this case is  accumulated in the 
double distributions
$F(x,y)$.  I studied the  evolution 
properties of  $F(x,y)$  and also showed that  $F(x,y)$'s are related
to the standard  distribution functions $f(x)$
through the reduction formulas (\ref{eq:redf}).
An interesting problem for the  future studies is to  understand
 a possible  interplay between the 
$x$- and $y$-dependencies of the double distributions
and construct 
phenomenological models for $F(x,y)$ . The simplest idea is to try 
a factorized ansatz  $F(x,y)= f(x) g(y)$.  However, 
the explicit expression for the evolution kernel 
$R(x,y;\xi,\eta)$  (which satisfies a similar reduction
formula (\ref{eq:rtop}) )  suggests that the structure
of  $F(x,y)$  is not necessarily as simple as that.

In the experimental aspect, 
as emphasized by Ji \cite{ji}, a continuous electron beam accelerator
with an energy 15 - 30 $GeV$ (like proposed ELFE)  may be  an ideal
place to study the scaling limit of the DVCS.
It  is also worth studying the possibility
of  observing   the first signatures of the 
scaling behavior of DVCS at CEBAF, especially at upgraded 
energies. 

In a forthcoming  paper \cite{gluon}, I discuss
the gluonic double distributions $F_g(x,y)$ 
which play a crucial role
in the  perturbative QCD approach to  hard diffractive 
electroproduction processes like 
\mbox{$\gamma^* p  \to p' \rho$.}

{\it Acknowledgments.}
I thank Xiangdong Ji for the discussion which stimulated 
this investigation and Charles  Hyde-Wright for persisting efforts
to  make me an enthusiast of  the virtual Compton scattering 
studies.
My special gratitude is to Nathan  Isgur for continued 
encouragement and advice.
This work is supported
by the US Department of Energy 
under contract DE-AC05-84ER40150.
I thank the DOE's Institute for Nuclear Theory
at the University of Washington for its hospitality
and support  during the workshop
``Quark and Gluon  Structure of Nucleons and Nuclei'' 
where this work has been started.


\begin{thebibliography}{99}

\bibitem{ji} X. Ji, preprint MIT-CTP-2517, Cambridge
 (1996); hep-ph/9603249.
\bibitem{gl}  V.N. Gribov and L.N. Lipatov, {\it Sov. J. Nucl. Phys. }
{\bf 15} (1972) 78; \\ L.N. Lipatov, {\it Sov. J. Nucl. Phys. } 
{\bf 20} (1975)  94.
\bibitem{ap}  G. Altarelli and G. Parisi, {\it  Nucl. Phys. }
{\bf B126} (1977) 298.
\bibitem{d}  Yu. L. Dokshitser, {\it Sov.Phys. JETP}, 
{\bf 46} (1977) 641.
\bibitem{bl}  S.J. Brodsky and G.P. Lepage,  {\it Phys.Lett.} 
{\bf 87B} (1979) 359.
\bibitem{tmf} A.V. Efremov and A.V. Radyushkin,  JINR report E2-11983,
Dubna (October 1978), \\ published in 
{\it Theor. Math. Phys.} {\bf 42} (1980) 97.
\bibitem{phlet} A.V. Efremov and A.V. Radyushkin,   {\it Phys.Lett.} 
 {\bf B94} (1980) 245.
\bibitem{spectral}  A.V. Radyushkin,  {\it Phys.Lett.}   
{\bf B131} (1983) 179.
\bibitem{bb} I.I.Balitsky and V.M.Braun, {\it Nucl.Phys. } 
{\bf B311} (1988/89) 541.
\bibitem{mikhrad} S.V. Mikhailov and A.V. Radyushkin, 
{\it  Nucl. Phys.} {\bf  B273} (1986) 297. 
\bibitem{gw} D.J. Gross and F. Wilczek,  {\it Phys. Rev.} {\bf D9}
(1974) 980.
\bibitem{gp}  H. Georgi and H.D. Politzer, {\it Phys. Rev.} {\bf D9} (1974) 416.
\bibitem{rr} A.V. Radyushkin and R. Ruskov, preprint 
CEBAF-TH-95-17, hep/ph 9511270 (to appear in {\it Phys.Lett. B}).
\bibitem{AuCh81} F.Del Aguila and M.K.Chase, {\it Nucl.Phys. } 
{\bf B193} (1981) 517.
\bibitem{braaten} E. Braaten,  {\it Phys. Rev.} {\bf D28 } (1983) 524.
\bibitem{kmr} E.P.Kadantseva, S.V.Mikhailov and A.V.Radyushkin,
     {\it Sov.J. Nucl.Phys.}  {\bf 44} (1986) 326.
\bibitem{gluon} A.V. Radyushkin (in preparation).  


\end{thebibliography}
\end{document}